\documentclass[twocolumn,prb,amsmath,amssymb,floatfix,superscriptaddress]{revtex4-2} 
\usepackage{amsmath}
\usepackage{amssymb}
\usepackage{graphicx}
\usepackage{bm}
\usepackage{graphicx}
\usepackage{upgreek}
\usepackage{scalerel}
\usepackage{multirow}
\usepackage{amsfonts}
\usepackage{color}
\usepackage{ulem}
\usepackage[breaklinks,hypertexnames=false]{hyperref}
\hypersetup{colorlinks,linkcolor={magenta},citecolor={blue},urlcolor={blue}}

 \usepackage[bb=dsserif]{mathalpha}
 
\usepackage{pgffor}
\usepackage{pdfpages}
\usepackage{mathdots}
\usepackage{float}
\usepackage{silence}
\usepackage{physics} 
\usepackage{lscape}   
\usepackage{color, colortbl}
\usepackage[table]{xcolor}
\usepackage{comment}  
\makeatletter
\AtBeginDocument{\let\LS@rot\@undefined}
\makeatother

\begin{document}

\title{Josephson effect and odd-frequency pairing in superconducting junctions with unconventional magnets}

\author{Yuri Fukaya}
\affiliation{Faculty of Environmental Life, Natural Science and Technology, Okayama University, 700-8530 Okayama, Japan}

\author{Kazuki Maeda}
\affiliation{Department of Applied Physics, Nagoya University, 464-8603 Nagoya, Japan}

\author{Keiji Yada}
\affiliation{Department of Applied Physics, Nagoya University, 464-8603 Nagoya, Japan}

\author{Jorge Cayao}
\affiliation{Department of Physics and Astronomy, Uppsala University, Box 516, S-751 20 Uppsala, Sweden}

\author{Yukio Tanaka}
\affiliation{Department of Applied Physics, Nagoya University, 464-8603 Nagoya, Japan}

\author{Bo Lu}
\affiliation{Center for Joint Quantum Studies, Tianjin Key Laboratory of Low Dimensional Materials Physics and Preparing Technology, Department of Physics, Tianjin University, Tianjin 300354, China}

\date{\today}

\begin{abstract}
We consider Josephson junctions formed by coupling two conventional superconductors via an unconventional magnet and investigate the formation of Andreev bound states, their impact on the Josephson effect, and the emergent superconducting correlations. 
In particular, we focus on unconventional magnets known as $d$-wave altermagnets and $p$-wave magnets. 
We find that the Andreev bound states in $d$-wave altermagnet and $p_y$-wave magnet Josephson junctions strongly depend on the transverse momentum, with a spin splitting and low-energy minima as a function of the superconducting phase difference $\varphi$. In contrast, the Andreev bound states for $p_{x}$-wave magnets are insensitive to the transverse momentum. We then show that the Andreev bound states can be probed by the local density of states in the middle of the junction, which also reveals that $d_{x^{2}-y^{2}}$- and $p$-wave magnet junctions are prone to host zero energy peaks. 
While the zero-energy peak in $d_{x^{2}-y^{2}}$-wave altermagnet junctions tends to oscillate with the magnetic order, it remains robust in $p$-wave magnet junctions.
We then discover that the Josephson current in $d$-wave altermagnet junctions is composed of higher harmonics of $\varphi$, which originate a $\phi$-Josephson junction behavior  entirely controlled by the magnetic order in $d_{xy}$-wave altermagnets.
In contrast, the Josephson current in Josephson junctions with $p$-wave magnets exhibits a conventional sine-like profile with a fast sign change at $\varphi=\pi$ due to zero-energy Andreev bound states. 
We also demonstrate that the critical currents in $d$-wave altermagnet Josephson junctions exhibit an oscillatory decay with the increase of the magnetic order, while the oscillations are absent in $p$-wave magnet junctions albeit the currents exhibit a slow decay. 
Furthermore, we also demonstrate that the interplay of the Josephson effect and unconventional magnetic order of $d$-wave altermagnets and $p$-wave magnets originates from odd-frequency spin-triplet $s$-wave superconducting correlations that are otherwise absent. 
Our results can serve as a guide to pursue the new functionality of  Josephson junctions based on unconventional magnets.
\end{abstract}

\maketitle

\section{Introduction}

The search for magnetic materials has attracted enormous interest in physics not only due to their fundamental properties but also due to their technological applications~\cite{Sarma_RevModPhys_2004,Eschrig2015,Baltz_RevModPhys_2018,HIROHATA2020166711,Newhorizons_spintronics}. 
Very recently, magnets possessing anisotropic spin-polarized Fermi surfaces in momentum space have been discovered and shown to exhibit zero net magnetization and energy bands with an intriguing spin-momentum locking~\cite{Bai_review24}. 
The most studied unconventional magnets have been shown to have $d$- and $p$-wave magnetic orders, and often known as altermagnets (AMs)~\cite{NakaNatCommun2019,Hayami19,NakaPRB2020,Hayami20,LiborSAv,MazinPNAS,Libor011028,MazinPRX22,landscape22,LiborPRX22} and unconventional $p$-wave magnets (UPMs)~\cite{hellenes2024P}, respectively. 
While both AMs and UPMs have a similar nonrelativistic spin splitting, and hence larger than relativistic spin-orbit coupled materials, such a nonrelativistic spin effect in both cases originates from distinct symmetries. 
On one hand, AMs are collinear-compensated magnets in real space that break time-reversal symmetry but preserve inversion symmetry; here opposite spins are connected by crystal rotation or mirror symmetries. 
AMs have already been found in various types of materials, such as in V$_2$Se$_2$O and V$_2$Te$_2$O~\cite{MaNatcommun2021,fzhang2024,jiang2024,hu2024}, metallic RuO$_{2}$~\cite{Ahn19,LiborSAv,LiborPRX22}, Mn$_{5}$Si$_{3}$~\cite{Helena2021,han2024SciAdv}, semiconducting-insulating La$_{2}$CuO$_{4}$~\cite{Moreno16}, MnTe~\cite{Lee24,Osumi2024,krempasky2024}, etc., see Ref.~\cite{Bai_review24} 
On the other hand, UPMs are noncollinear and noncoplanar magnets preserving time-reversal symmetry but breaking inversion symmetry; UPMs are protected by the combination of time-reversal and a translation of a half the unit cell~\cite{hellenes2024P}.  
Candidate materials for UPMs are Mn$_{3}$GaN and CeNiAsO~\cite{hellenes2024P}.  Thus, the enormous recent advances in unconventional magnets show that it is timely to exploit their potential for realizing even more exotic phases.

One of the possible routes is to combine unconventional magnets with superconductors (SCs), which is not only important for understanding superconductivity in AMs and UPMs, but also for applications in superconducting spintronics \cite{Eschrig2015,kokkeler2024}. 
For AM-SC and UPM-SC junctions, it has been shown that the Andreev reflection strongly depends on the crystal orientation and the strength of the spin-splitting~\cite{Sun23,Papaj23,maeda2024,nagae2024,Bo2025}. 
In Josephson junctions with an AM and conventional spin-singlet $s$-wave SCs, the ground state has been shown to exhibit $0$-$\pi$ oscillations~\cite{Ouassou23,Beenakker23,zhang2024} but also free energy minima away from $0,\pi$ with double degeneracy~\cite{Bo2024}, see also Refs.~\cite{tanaka961,tanaka971,Buzdin03,Goldovin,Golubov2004}. 
The $0$-$\pi$ transitions have also been shown to persist in altermagnet-based Josephson junctions between spin-singlet and spin-triplet SCs~\cite{Cheng24}, albeit with features of spin-triplet superconductivity.  
The current-phase curves in altermagnet-based Josephson junctions develop anomalous features, including multiple nodes~\cite{Bo2024}, tunable skewness~\cite{sun2024}, and multiple harmonics of the superconducting phase difference. 
Furthermore, the interplay of AMs and SCs has also been shown to lead to exotic phenomena, such as topological superconductivity~\cite{Zhongbo23,CCLiu1,CCLiu2,Brekke23,Cano23}, diode effect~\cite{Banerjee24,Qiang24,chakrabortyd2024}, and magnetoelectric effect~\cite{zyuzin2024,huN2024}.

Despite the intense efforts in altermagnet-based Josephson junctions, there are still open questions.  For instance, it is poorly understood what is the role of Andreev bound states (ABSs) even though they are expected to carry the supercurrent across the junction \cite{Kashiwaya_2000,sauls2018andreev,mizushima2018multifaceted}.  
Moreover, since the Josephson effect involves the transfer of Cooper pairs between superconductors \cite{mahan2013many,zagoskin}, it is expected that the Josephson effect reveals the type of superconducting pairing \cite{Tanaka2012,Cayao2020} but it has largely remained unexplored.  Another problem is that most of the previous altermagnet-based Josephson studies consider continuum models to describe AMs, which turn out to be challenging due to the anisotropic Fermi surface of AMs. 
Even more intriguing, most of the previous research has focused on Josephson junctions based on AMs but the Josephson effect in junctions based on UPMs has not been addressed yet.  Therefore, there are still several open problems in Josephson junctions based on unconventional magnets.

In this work, we consider Josephson junctions based on spin-singlet $s$-wave superconductors and unconventional magnets, with the aim to study the emergence of Andreev bound states and their role on the phase-biased Josephson transport as well as on understanding proximity-induced superconducting correlations. Particularly,  we focus on unconventional magnets with $d$-wave and $p$-wave magnetic order, referred to in the previous paragraphs as AMs and UPMs, respectively. We find that the emergent Andreev bound states in Josephson junctions with $d$-wave AMs and $p_y$-wave UPM strongly depend on the transverse momentum, while for $p_x$-wave UPM   they are insensitive. We then show that the Andreev bound states can be detected by the local density of states (LDOS), which unveils the presence of zero-energy peaks in $d_{x^{2}-y^{2}}$- and $p$-wave magnet junctions entirely determined by the magnetic order. In fact, we obtain that only the zero-energy LDOS peaks in $d_{x^{2}-y^{2}}$-wave AM junctions oscillate with the magnetic order, with the periodicity of oscillations directly related to the magnetic field strength. 

We then demonstrate that the phase-biased Josephson currents exhibit a nontrivial phase dependence in the case of $d$-wave junctions, with contributions from higher harmonics of $\varphi$ that enable the realization of $\phi$-Josephson junctions with $d_{xy}$-wave AMs. Interestingly, the critical currents in $d$-wave altermagnet junctions develop an oscillatory decaying profile due to the magnetic order, while the current in $p$-wave UPM junctions only exhibits a slow decay without oscillations. Moreover, we prove that the interplay of inherent unconventional magnetism and the Josephson effect permits the realization of proximity-induced odd-frequency spin-triplet superconducting pairing. Our results provide a fundamental understanding of superconductivity in Josephson junctions with unconventional magnets, which can be useful for their future application in superconducting spintronics. 

The remainder of this work is organized as follows. 
In  Sect.~\ref{section2} we 
present the model of the considered Josephson junction with unconventional magnets and conventional spin-singlet superconductors. 
Here, we also discuss the method to obtain the formation of ABSs, LDOS, Josephson currents, and superconducting correlations. 
In Sect.~\ref{section3}, we discuss the formation of ABSs and their impact on the LDOS in the middle of the junction. 
In Sect.~\ref{section4}, we present the current-phase curves and corresponding free energies, while in Sect.~\ref{section5} we discuss the emergent odd-frequency superconducting correlations.  
In Sect.~\ref{section6}, we present our conclusions.

\section{Modelling Josephson junctions with unconventional magnets}
\label{section2}
We focus on Josephson junctions formed by conventional spin-singlet $s$-wave SCs and an unconventional magnet that can be an AM or UPM, see Fig.~\ref{fig:1}(a). In part, we discuss the Hamiltonian models we consider for the SCs and unconventional magnets. Moreover, here, we also outline how the method to obtain the ABSs, LDOS, Josephson current, and superconducting correlations.

\subsection{Bulk Hamiltonian models}
\begin{figure}[t!]
    \centering
    \includegraphics[width=8.5cm]{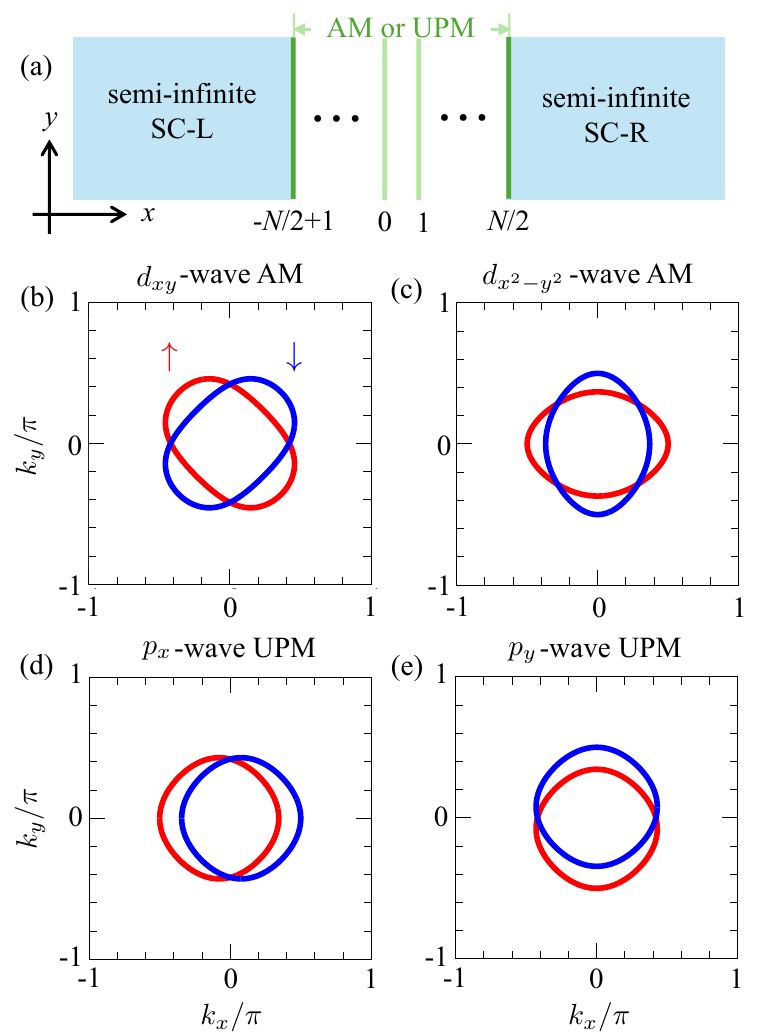}
    \caption{(a) Illustration of a two-dimensional Josephson junction along the $x$-direction formed by semi-infinite SCs and a finite length unconventional magnet (AM or UPM). The length of the unconventional magnet is  $L=Na$, with $N$ being the number of lattice sites and $a$ the lattice spacing.  
    We consider the periodic boundaries along  $y$, so that $k_{y}$ is a good quantum number. (b) and (c) show the Fermi surface for AMs with $d_{xy}$- and $d_{x^2-y^2}$-wave magnetic order, while (d) and (e) for UPM with $p_{x}$- and $p_{y}$-wave. The red and blue curves in (b)-(e) indicate the up and down spins, respectively. Parameters for (b-e):   $\alpha_{1}=0.5t$, $\alpha_{2}=0.5t$,  $\beta_{1}=0.5t$, $\beta_{2}=0.5t$, and $\mu=1.5t$.}
    \label{fig:1}
\end{figure}

The unconventional magnet is in the normal state (N) and modeled by 
\begin{equation}
\label{EqNAM}
    \hat{H}_{\rm N}(\bm{k})=\hat{H}_{0}(\bm{k})+\hat{H}_\mathrm{M}(\bm{k}).
\end{equation}%
where
\begin{equation}
\label{energy1}
\begin{split}
     \hat{H}_{0}(\bm{k})&=[-\mu+4t_{1}-2t_{1}\cos{k_x}-2t_{1}\cos{k_y}]\hat{s}_{0},\\   
     \hat{H}_\mathrm{M}(\bm{k})&=[2\alpha_1\sin{k_x}\sin{k_y}+\alpha_2(\cos{k_x}-\cos{k_y})\\
    &+\beta_1\sin{k_x}+\beta_2\sin{k_y}]\hat{s}_{z},  
\end{split}
\end{equation}
describe the kinetic energy and the unconventional magnet, respectively. 
Here, $\bm{k}=(k_{x},k_{y})$ is the two-dimensional wavevector, $\hat{s}_{0,x,y,z}$ represent the Pauli matrices in spin space, $\mu=1.5t$ is the chemical potential, and $t_{1}=t$ is the hopping integral. 
Moreover, $\alpha_{1}$ and $\alpha_{2}$ represent the amplitude of $d_{xy}$- and  $d_{x^2-y^2}$-wave AMs, while $\beta_{1}$ and $\beta_{2}$ represent the amplitude of $p_x$- and $p_y$-wave magnets, respectively. 
Here, the momentum dependence of the exchange field in UPM is different from the common Dirac-like spin-orbit coupling~\cite{maeda2024} but it is equivalent to the momentum profile in a persistent helix~\cite{BerevigPRL2006,KohdaPRB2012,LiuPRL2014,IkegayaPRB2015,JacobsenPRB2015,YangPRB2017,AlidoustPRB2020,IkegayaPRBL2021,AlidoustPRBL2021,LeePRB2021}. We however note that the UPM order is generated due to the phase transition in contrast to what occurs in the persistent helix~\cite{BerevigPRL2006,KohdaPRB2012,LiuPRL2014,IkegayaPRB2015,JacobsenPRB2015,YangPRB2017,AlidoustPRB2020,IkegayaPRBL2021,AlidoustPRBL2021,LeePRB2021}. To understand the features of the unconventional magnet, in Figs.~\ref{fig:1}(b) \ref{fig:1}(e), we plot their Fermi surfaces by using  Eq.\ (\ref{EqNAM}) and (\ref{energy1}). In the presence of an anisotropic exchange field, the spin degeneracy of the Fermi surface is lifted in both AMs and UPMs, as shown in Figs.~\ref{fig:1} (b) \ref{fig:1}(e). Moreover, the Fermi surfaces for distinct spins depend on momenta, signaling their inherent anisotropic nature of AMs and UPMs. 
While in $d$-wave AMs the Fermi surfaces for each spin depend differently on both $k_{x}$ and $k_{y}$, $p_{x(y)}$-wave magnets exhibit a shift of their Fermi surfaces along $x (y)$.
We stress that the anisotropic Fermi surfaces for AMs and UPMs are distinct from the circle-type Fermi surface of normal metals.

The conventional SCs are modeled with a Bogoliubov-de Gennes Hamiltonian given by
\begin{align}
    \hat{H}_\mathrm{SC}(\bm{k})=
        \begin{pmatrix}
        \hat{H}_{0}(\bm{k}) & \hat{\Delta}(\bm{k})\\
        \hat{\Delta}^{\dagger}(\bm{k}) & -\hat{H}^{*}_{0}(-\bm{k})
    \end{pmatrix}, \label{BdG_k}
\end{align}%
where 
\begin{equation}
\label{pair_potential}
\hat\Delta(\bm{k})=i\sigma_{y}\Delta
\end{equation}
is the spin-singlet $s$-wave pair potential. Based on the BCS theory \cite{tinkham2004introduction}, the pair potential obeys $|\Delta|=\Delta_{0}\tanh[1.74\sqrt{(T_\mathrm{c}-T)/T}]$, where $ \Delta_{0}=3.53T_\mathrm{c}/2$; here, $T$ is the temperature while $T_\mathrm{c}$ the critical temperature chosen as $T_{c}=0.01t$. 
Since we are interested in   Josephson junctions based on unconventional magnets and SCs [Fig.\,\ref{fig:1}(a)], in what follows we discuss how we model them based on Eq.\,(\ref{EqNAM}) and Eq.\,(\ref{BdG_k}).
  
\subsection{Josephson junction model}
We consider a Josephson junction along the $x$-direction formed by two conventional $s$-wave SCs modeled by Eqs.\ (\ref{BdG_k}) mediated by a finite length unconventional magnet given by Eq.~\ref{EqNAM}. 
Along $y$ we consider periodic boundary conditions.  
For this reason, we discretize the momentum along $x$ and take the middle region to be of length $L=Na$, where $N$ is the number of the lattice sites and $a$ is the lattice spacing set to $a=1$ in this work. 
Thus, the unconventional magnet Hamiltonian given by Eq.\,(\ref{EqNAM}) is now written in real space $H_{\rm N}(k_{y})=\sum_{j}\hat{u}_\mathrm{N}(k_y)c^{\dagger}_{j}c_{j}+\hat{t}_\mathrm{N}(k_{y})c^{\dagger}_{j}c_{j+1}+{\rm h.\,c.}$, where $\hat{u}_\mathrm{N}(k_y)$ and $\hat{t}_\mathrm{N}(k_{y})$ are onsite energies and nearest-neighbor hopping, while $j$ runs over the N lattice sites inside the unconventional magnet, namely, $j\in[-N/2+1,N/2]$, see Fig.\,\ref{fig:1}(a). 
In BdG form, $\hat{u}_\mathrm{N}(k_y)$ and $\hat{t}_\mathrm{N}(k_{y})$ are given by
\begin{equation}
\label{uNtN}
\begin{split}
    \hat{u}_\mathrm{N}(k_y)&=
    \begin{pmatrix}
        \tilde{u}_\mathrm{N}(k_y) & 0 \\
        0 & -\tilde{u}^{*}_\mathrm{N}(-k_y)
    \end{pmatrix}\,,\\
     \hat{t}_\mathrm{N}(k_{y})&=
    \begin{pmatrix}
        \tilde{t}_\mathrm{N}(k_y) & 0 \\
        0 & -\tilde{t}^{*}_\mathrm{N}(-k_y)
    \end{pmatrix},\\
\end{split}
\end{equation}
with 
\begin{equation}
\label{tildeuNtN}
\begin{split}
  \tilde{u}_\mathrm{N}(k_y)&=[-\mu+4t_1-2t_1\cos{k_y}]\hat{s}_{0}\\
&    +[-\alpha_{2}\cos{k_y}+\beta_{2}\sin{k_y}]\hat{s}_{z}\,,\\
     \tilde{t}_\mathrm{N}(k_y)&=-t_{1}\hat{s}_{0}+\left[-i\alpha_{1}\sin{k_y}+\frac{\alpha_{2}}{2}-i\frac{\beta_{1}}{2}\right]\hat{s}_{z}\,.
    \end{split} 
    \end{equation}
Here, $\alpha_{1,2}$ and $\beta_{1,2}$ describe the $d$-wave AMs and the unconventional $p$-wave magnets, respectively. 
For $\alpha_{1,2}=0$ and $\beta_{1,2}=0$, Eqs.\,(\ref{uNtN}) and (\ref{tildeuNtN}) describe the onsite and nearest-neighbor hopping in the normal state of the SCs resulting from $ \hat{H}_{0}$ in Eq.\,(\ref{BdG_k}). 

When it comes to the pair potential, we note that the middle region is in the normal state and only the left and right SCs exhibit a finite pair potential. For this reason, the pair potential throughout the junction is given by
 \begin{equation}
 \label{DeltaSNS}
    \hat{\bf\Delta}(x)=
    \begin{cases}
        \hat{\Delta}\,, & x< -N/2+1\,,\\
        0\,, & -N/2+1\leq x\leq N/2\,,\\
        \hat{\Delta}e^{-i\varphi}\,, & x> N/2\,,
    \end{cases}
\end{equation}%
where $\varphi$ is the phase difference between SCs and $\hat{\Delta}$ the pair potential in the SCs having the same matrix structure as in Eq.\ (\ref{pair_potential}).  We note that here we consider that the SCs are semi-infinite, as sketched in Fig.\,\ref{fig:1}. 
Thus, combining Eqs.\,(\ref{uNtN}), together with Eqs.\,(\ref{tildeuNtN}) and Eqs.\,(\ref{DeltaSNS}) we model a Josephson junctions with two semi-infinite SCs and an unconventional magnet of length $L$.

\subsection{Obtaining the ABSs, LDOS, Josephson current, and pair amplitudes}
A finite difference between SCs enables the formation of ABSs \cite{sauls2018andreev}, which then gives rise to the phase-dependent supercurrent characterizing the Josephson effect \cite{Golubov2004}. 
In this process, the superconducting correlations are involved because the supercurrent is based on the transfer of Cooper pairs between SCs \cite{mahan2013many,zagoskin}. 
These reasons motivate us to explore the impact of the ABSs on the Josephson effect as well as on the emergent superconducting correlations. 
For completeness, we also inspect how ABSs can be detected via LDOS.  
More specifically, by focusing on the two middle sites $j=0,1$, the LDOS and supercurrent are calculated as~\cite{YadaJPSJ2014}
\begin{equation}
\label{LDOSI}
\begin{split}
    D(E,\varphi)&=-\frac{1}{\pi}\int^{\pi}_{-\pi}\mathrm{Tr'}[\mathrm{Im}[\hat{G}_{00}(k_y,E)]]dk_y\,,\\
    I(\varphi)&=\frac{iet_{1}}{\hbar}
    \int^{\pi}_{-\pi}\mathrm{Tr'}T\notag\\
    &\times\sum_{i\varepsilon_{n}}[\hat{G}_{01}(k_y,i\varepsilon_n)-\hat{G}_{10}(k_y,i\varepsilon_n)]dk_y\,,
\end{split}
\end{equation}
where $\hat{G}_{jj'}$ is the Green's function in Nambu space in the middle of the unconventional magnet with $j,j'=0,1$: $\hat{G}_{00}$ and $\hat{G}_{11}$ are the local Green's function at sites $j=0,1$, while  $\hat{G}_{01}$ and $\hat{G}_{10}$ represent the nonlocal Green's function between middle sites $j=0,1$. Moreover, $\mathrm{Tr'}$ means that the trace is taken only over the electron subspace, $i\varepsilon_n=i(2n+1)\pi T$ are Matsubara frequencies, and $T$ is the temperature, while $E$ represents real energies and $\delta$ is an infinitesimal small positive number. In practice, $\hat{G}_{jj'}$ is obtained in terms of Matsubara frequencies but for the LDOS we perform the well-known analytic continuation $i\varepsilon\rightarrow E+i\delta$ so that we get the retarded Green's function. The Greens functions  $\hat{G}_{jj'}$ in the middle of the unconventional magnet are calculated as~\cite{Umerski97,KawaiPRB2017,FukayaPRB2020,Fukayanpj2022} 
\begin{equation}
\label{g00g11}
\begin{split}
    \hat{G}_{00}(k_y,i\varepsilon_n)
    &=[[\hat{G}^{(0)}_\mathrm{L}(k_y,i\varepsilon_n)]^{-1}-\hat{t}_\mathrm{N}\hat{G}^{(1)}_\mathrm{R}(k_y,i\varepsilon_n)\hat{t}^{\dagger}_\mathrm{N}]^{-1},\\
    \hat{G}_{11}(k_y,i\varepsilon_n)
    &=[[\hat{G}^{(1)}_\mathrm{R}(k_y,i\varepsilon_n)]^{-1}-\hat{t}^{\dagger}_\mathrm{N}\hat{G}^{(0)}_\mathrm{L}(k_y,i\varepsilon_n)\hat{t}_\mathrm{N}]^{-1}, 
    \end{split}
\end{equation}
for the local components, while for the nonlocal parts we use
\begin{equation}
\label{g01g10}
\begin{split}
   \hat{G}_{01}(k_y,i\varepsilon_n)&=\hat{G}^{(0)}_\mathrm{L}(k_y,i\varepsilon_n)\hat{t}_\mathrm{N}\hat{G}_{11}(k_y,i\varepsilon_n),\\
    \hat{G}_{10}(k_y,i\varepsilon_n)&=\hat{G}^{(1)}_\mathrm{R}(k_y,i\varepsilon_n)\hat{t}^{\dagger}_\mathrm{N}\hat{G}_{00}(k_y,i\varepsilon_n)\,. 
    \end{split}
\end{equation}%
Here, $\hat{G}^{(0)}_\mathrm{L}$, $\hat{G}^{(1)}_\mathrm{R}$ are the surface Green's functions of each side of the normal regions of the junctions and are obtained by
\begin{align}
    \hat{G}^{(j)}_\mathrm{L}(k_y,i\varepsilon_n)
    &=[i\varepsilon_n-\hat{u}_\mathrm{N}(k_y)-\hat{t}^{\dagger}_\mathrm{N}\hat{G}^{(j-1)}_\mathrm{L}(k_y,i\varepsilon_n)\hat{t}_\mathrm{N}]^{-1}, \label{GLj}\\
    \hat{G}^{(j)}_\mathrm{R}(k_y,i\varepsilon_n)
    &=[i\varepsilon_n-\hat{u}_\mathrm{N}
    (k_y)-\hat{t}_\mathrm{N}\hat{G}^{(j+1)}_\mathrm{R}(k_y,i\varepsilon_n)\hat{t}^{\dagger}_\mathrm{N}]^{-1}.
    \label{GRj}
\end{align}%
where $\hat{G}^{(j\mp 1)}_\mathrm{L,R}$ in the left and right-hand sides are calculated by Eqs.\ (\ref{GLj}) and (\ref{GRj}) recursively~\cite{Umerski97,KawaiPRB2017,FukayaPRB2020,Fukayanpj2022}.
The superconductors are incorporated via surface Green's functions as
\begin{align}
    \hat{G}^{(-N/2+1)}_\mathrm{L}(k_y,i\varepsilon_n)
    &=[i\varepsilon_n-\hat{u}_\mathrm{N}(k_y)-\hat{t}^{\dagger}_\mathrm{J}\hat{G}_\mathrm{L}(k_y,i\varepsilon_n)\hat{t}_\mathrm{J}]^{-1},\\
    \hat{G}^{(N/2)}_\mathrm{R}(k_y,i\varepsilon_n)
    &=[i\varepsilon_n-\hat{u}_\mathrm{N}(k_y)-\hat{t}_\mathrm{J}\hat{G}_\mathrm{R}(k_y,i\varepsilon_n)\hat{t}^{\dagger}_\mathrm{J}]^{-1},
    \label{G0110}
\end{align}%
where the superscript $j=(-N/2+1,N/2)$ is the site index inside the normal regions (AM or UPM).
Here, $\hat{t}_\mathrm{J}$ denotes the tunneling Hamiltonian:
\begin{align}
    \hat{t}_\mathrm{J}
    =t_\mathrm{int}
    \begin{pmatrix}
    -t_{1} & 0 \\
    0 & -t_{1}
    \end{pmatrix}
    \otimes\hat{\tau}_{3},
\end{align}%
with the transparency $t_\mathrm{int}=0.95$ and the Pauli matrices in Nambu space $\hat{\tau}_{i=0,x,y,z}$.
We obtain two semi-infinite Green's functions in the left-side $\hat{G}_\mathrm{L}(k_y,i\varepsilon_n)$ and the right-side superconductors
$\hat{G}_\mathrm{R}(k_y,i\varepsilon_n)$ by using the M\"{o}bius transformation as discussed in Refs.~\cite{Umerski97,KawaiPRB2017,FukayaPRB2020,Fukayanpj2022}. 

The Green's functions $\hat{G}_{jj'}$ in the middle of the unconventional magnet obtained by Eqs.\,(\ref{g00g11}) and (\ref{g01g10}) also allow the calculation of the emergent superconducting correlations. This is because, the Green's functions $\hat{G}_{jj'}$ are in Nambu space and has the following structure,
\begin{align}
\label{GF}
    \hat{G}_{j,j'}(k_y,i\varepsilon_{n})
    &=
    \begin{pmatrix}
        \tilde{G}_{j,j'}(k_y,i\varepsilon_{n}) & \hat{F}_{j,j'}(k_y,i\varepsilon_{n})\\
        \bar{F}_{j,j'}(k_y,i\varepsilon_{n}) & \bar{G}_{j,j'}(k_y,i\varepsilon_{n})
    \end{pmatrix}.
\end{align}%
where the diagonal entries ($\tilde{G}_{j,j'}$ and $\bar{G}_{j,j'}$) represent the normal (electron-electron and hole-hole) Green's functions, while the off-diagonal entries ($\hat{F}_{j,j'}$ and $\bar{F}_{j,j'}$) correspond to the anomalous (electron-hole and hole-electron) Green's functions. It is therefore evident that the anomalous Green's functions characterize the superconducting correlations and will be used here to identify the symmetry of the emergent superconducting pairing.

Therefore, by using Eqs.\,(\ref{LDOSI}), (\ref{G0110}), and (\ref{GF}) we numerically calculate the LDOS, Josephson currents, and emergent superconducting correlations. 
We stress that the Green's function in Eq.\ (\ref{GF}) also depends on the phase difference $\varphi$, implying that observables calculated using it will depend on $\varphi$.
Using the poles of the local Green's function $\hat{G}_{00}(k_y,E)$ we also obtain the formation of ABSs, which we discuss next. 
Unless otherwise specified, we consider unconventional magnets of length $L=30a$; moreover, we take a critical temperature of $T_{c}=0.01t$, a temperature of $T=0.025T_{c}$, implying a pair potential $\Delta=0.018t$. 
This corresponds to a superconducting coherence length of $\xi\sim100a$, which places the considered Josephson junctions within the short junction regime.


\begin{figure*}[htbp]
    \centering
    \includegraphics[width=16cm]{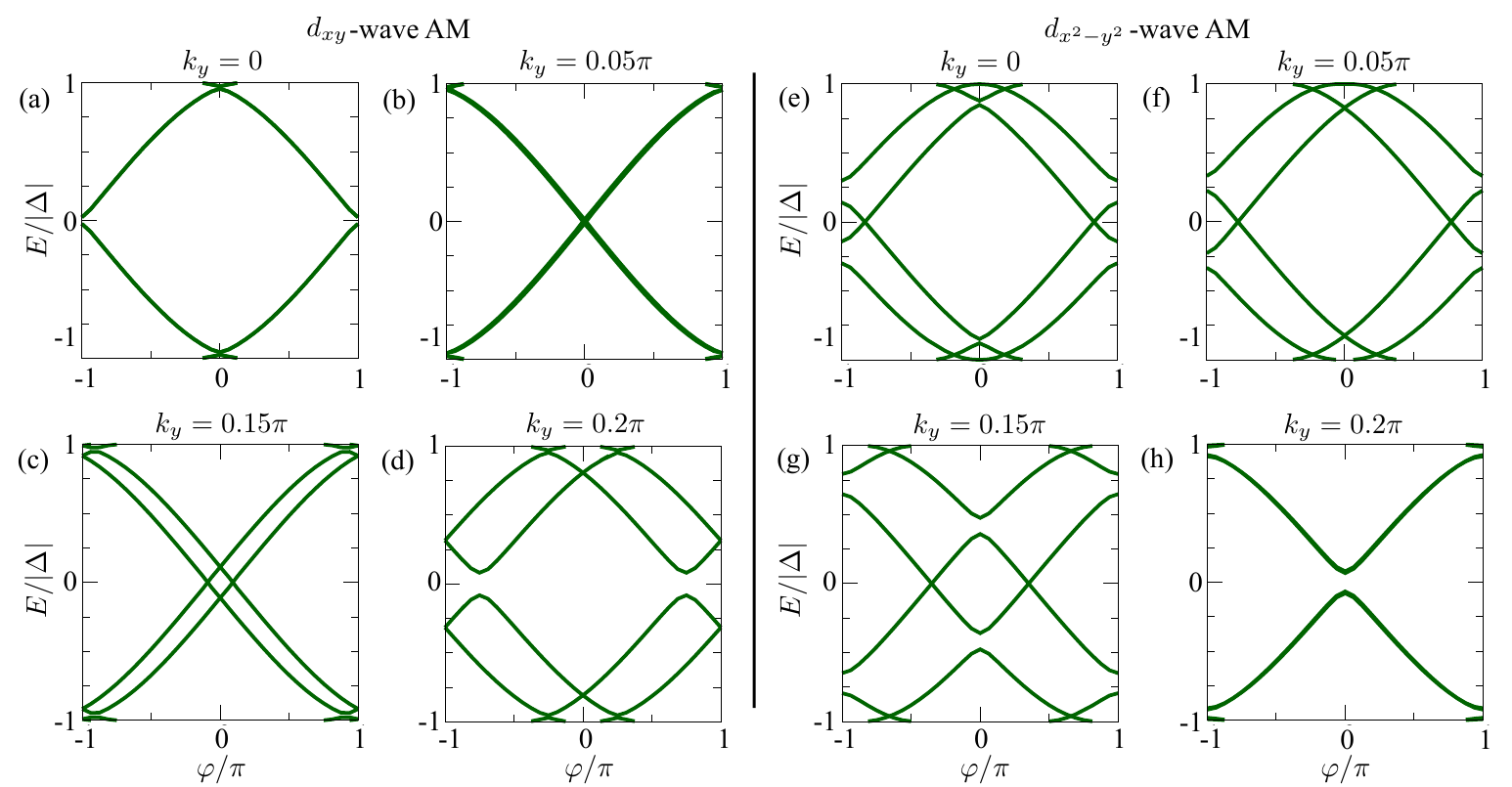}
    \caption{(a)-(d) Andreev bound states as a function of $\varphi$ for a $d_{xy}$-wave AM Josephson junction as a function of $\varphi$, where different values correspond to distinct values of the transverse momentum $k_{y}$. (e)-(h) The same as in (a)-(d) but for a $d_{x^{2}-y^{2}}$-wave AM Josephson junction. Parameters: $\alpha_{1}=0.35t$,  $\alpha_{2}=0.24t$, $\mu=1.5t$, $|\Delta|=0.018t$, $t_\mathrm{int}=0.95$, and $N=30$.}
    \label{fig:2}
\end{figure*}%

\begin{figure}[htbp]
    \centering
    \includegraphics[width=8.5cm]{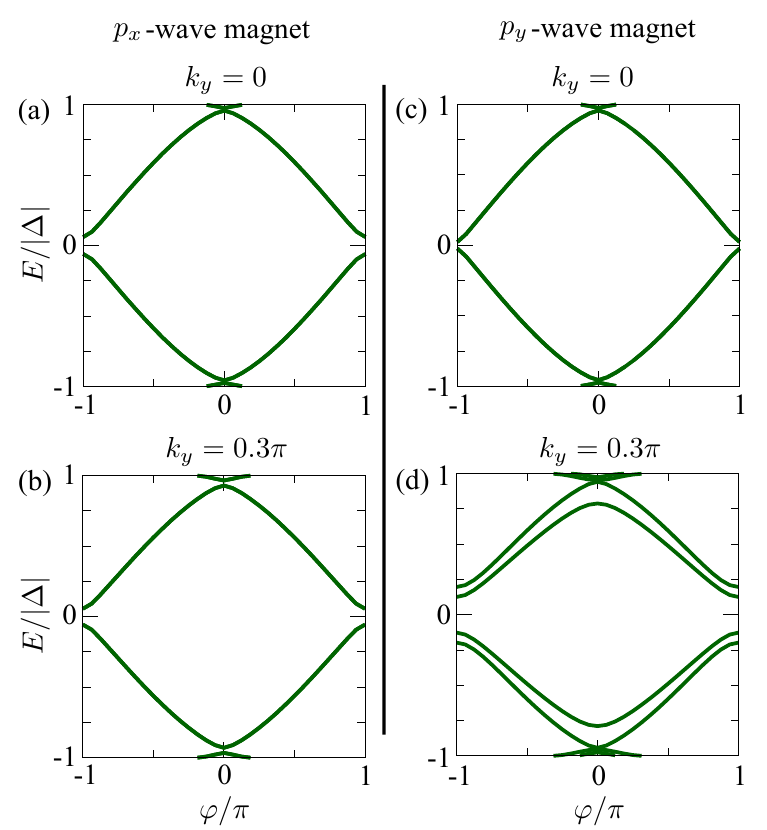}
    \caption{Andreev bound states for a $p_{x,y}$-wave magnet Josephson junction as a function of $\varphi$ at distinct values of the transverse momentum $k_{y}=0,0.3\pi$. 
Parameters:  $\beta_{1}=0.5t$, $\beta_{2}=0.5t$, and other parameters same  as in Fig.~\ref{fig:2}. 
    }
    \label{fig:3}
\end{figure}%

\section{Andreev bound states and LDOS}
\label{section3}
Here, we discuss the formation of ABSs and their signatures in the LDOS in the Josephson junctions presented in the previous section. 
We numerically obtain the ABSs from the poles of the local Green's function in the center of the unconventional magnet, $\hat{G}_{00}(k_y,E)$. 
In Fig.\,\ref{fig:2} we present the ABSs as a function of the superconducting phase difference $\varphi$ for distinct $k_{y}$ in a Josephson junction with a middle region being $d_{xy}$- and $d_{x^{2}-y^{2}}$-wave AMs. 
In Fig.\,\ref{fig:3} we show the ABSs as a function of $\varphi$ for a Josephson junction with a $p_{x(y)}$-wave magnet. 
Under general circumstances, ABSs appear in all cases strongly dependent on the phase difference but develop particular differences depending on the type of unconventional magnet. 

In the case of Josephson junctions with a $d_{xy}$-wave AM, there is a pair of spin degenerate ABSs at $k_{y}=0$ dispersing with $\varphi$ in a cosine-like fashion [Fig.\,\ref{fig:2}(a)], as expected for a Josephson junction with conventional SCs since the AM field here vanishes. 
A nonzero $k_{y}$ induces a finite AM field that strongly affects the phase dependent ABSs, see Fig.\,\ref{fig:2}(b-d). 
In fact, at $k_{y}=0.05\pi$ in Fig.\,\ref{fig:2}(b), the ABSs reach zero energy at $\varphi=0$, which signals a phase shift entirely due to the $d_{xy}$-wave AM.  
Furthermore, larger values of $k_{y}$ induce a visible spin splitting between ABSs which then tend to cross zero energy at phases around $\varphi=0$, see Fig.\,\ref{fig:2}(c). As $k_{y}$ further increments, the spin splitting takes larger values, and the ABSs close to zero energy tend to approach zero energy at phases close to $\varphi=\pm\pi$, see Fig.\,\ref{fig:2}(d). 

For Josephson junctions with a  $d_{x^{2}-y^{2}}$-wave AM, the ABSs exhibit a similar dependence on the AM field but with key differences, see Fig.\,\ref{fig:2}(e)-\ref{fig:2}(h). 
For instance, already at $k_{y}=0$ the ABSs are split in spin [Fig.\,\ref{fig:2}(e)], unlike $d_{xy}$-wave case. 
This is of course an expected result because at $k_{y}=0$, there is a finite strength of the $d_{x^{2}-y^{2}}$-wave altermagnet field, see Eq.\,(\ref{energy1}). 
Here, at $k_{y}=0$ in Fig.\,\ref{fig:2}(e), the spin-split ABSs closest to zero tend to cross zero energy at phases close to $\varphi=\pm\pi$, which then shifts to phases around $\varphi=0$ when $k_{y}$ increases [Fig.\,\ref{fig:2}(f)(g)]. 
Interestingly, at $k_{y}=0.2\pi$, the AM field moves the first excited pair of ABSs and leaves only a single pair of ABSs around zero energy, see Fig.\,\ref{fig:2}(h). 
This single pair of ABSs tends to reach zero energy at $\varphi=0$ [Fig.\,\ref{fig:2}(h)], revealing a clear phase shift with respect to the case with $k_{y}=0$ in Fig.\,\ref{fig:2}(e). 
The finite energy at $\varphi=0$ of the ABSs depends on the system parameters, such as the chemical potential; that is why also here zero-energy ABSs can be also obtained.
The sensitive dependence of ABSs on $k_y$ for $d$-wave AM-based Josephson junctions reflects the alternating spin-splitting field in AM, which is similar to ferromagnetic \cite{TANAKAFI} or topological Josephson junctions \cite{PhysRevB.91.024514,cayao2018andreev,Bo2023}.

When it comes to Josephson junctions with UPMs, the ABSs develop a cosine-like phase-dependent profile but with different spin splitting for $p_{x}$- and $p_{y}$-wave magnets, as seen in Fig.\,\ref{fig:3}. 
For $p_{x}$-wave-based junctions, the ABSs exhibit roughly the same phase dependence for distinct values of $k_{x}$ [Figs.\,\ref{fig:3}(a) and \ref{fig:3}(b)], which is expected because the AM field here behaves as a spin-orbit coupling field that only produces a slight shift on the wavevector along $x$. This behavior has also been found in Josephson junctions with spin-orbit coupling \cite{PhysRevB.91.024514,cayao2018andreev,cayao2018finite,PhysRevB.96.155426}; it is worth noting that the conventional cosine-like profile seen here is similar to what is expected for Josephson junctions with conventional SCs mediated by normal metals \cite{sauls2018andreev,mizushima2018multifaceted}. 
In the case of Josephson junctions with $p_{y}$-wave magnets, the ABSs undergo a clear spin splitting at finite $k_{y}$ because it acts here as a magnetic field along $y$, see Figs.\,\ref{fig:3}(c) and \ref{fig:3}(d). 
In spite of the spin splitting for $p_{y}$-wave magnets, the two pairs of spin-split ABSs have a cosine dependence with the phase difference, see Fig.\,\ref{fig:3}(d). 
This behavior is distinct to what we obtained for Josephson junctions with $d$-wave AMs in Fig.\,\ref{fig:2}.

 \begin{figure}[htbp]
    \centering
    \includegraphics[width=8.4cm]{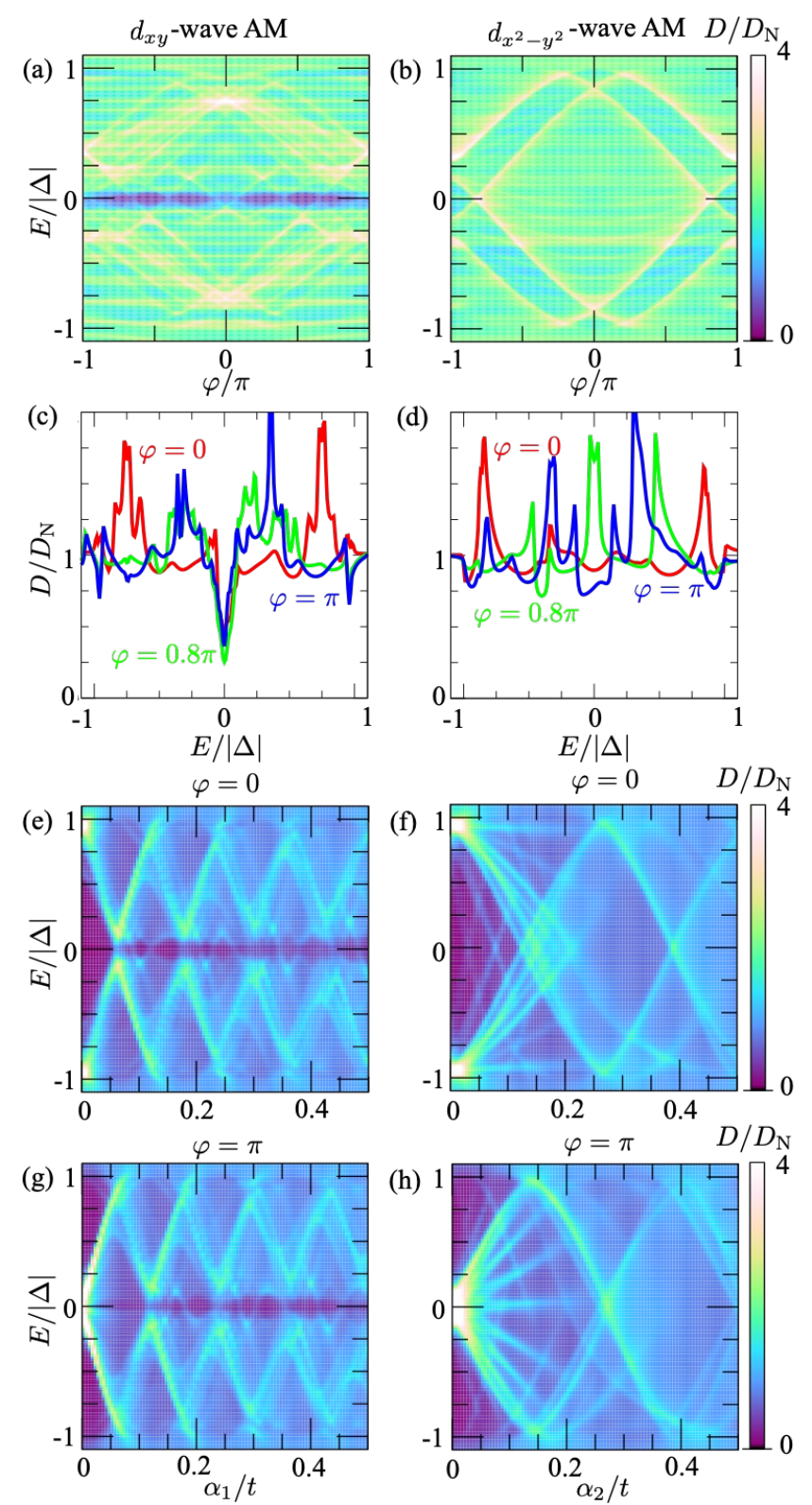}
    \caption{(a) and (b) LDOS normalized by its zero-energy normal state value as a function of $\varphi$ and $E$ in the middle of a Josephson junction with $d_{xy}$ and $d_{x^2-y^2}$-wave AMs. 
    (c) and (d) Line cuts of (a) and (b) as a function of $E$ at $\varphi=0,0.8\pi,\pi$. 
    (e) and (g) Normalized LDOS as a function of $E$ and strength of magnetic $d_{xy}$-wave order $\alpha_{1}$ at $\varphi=0,\pi$. (f) and (h) Normalized LDOS as a function of $E$ and strength of magnetic $d_{x^{2}-y^{2}}$-wave order $\alpha_{2}$ at $\varphi=0,\pi$. Parameters: $(j,j')=(0,0)$, $\alpha_{1}=0.35t$, $\alpha_{2}=0.24t$, $\mu=1.5t$, $|\Delta|=0.018t$, $t_\mathrm{int}=0.95$, and $N=30$.
    }
    \label{fig:4}
\end{figure}%
\begin{figure}[htbp]
    \centering
    \includegraphics[width=8.4cm]{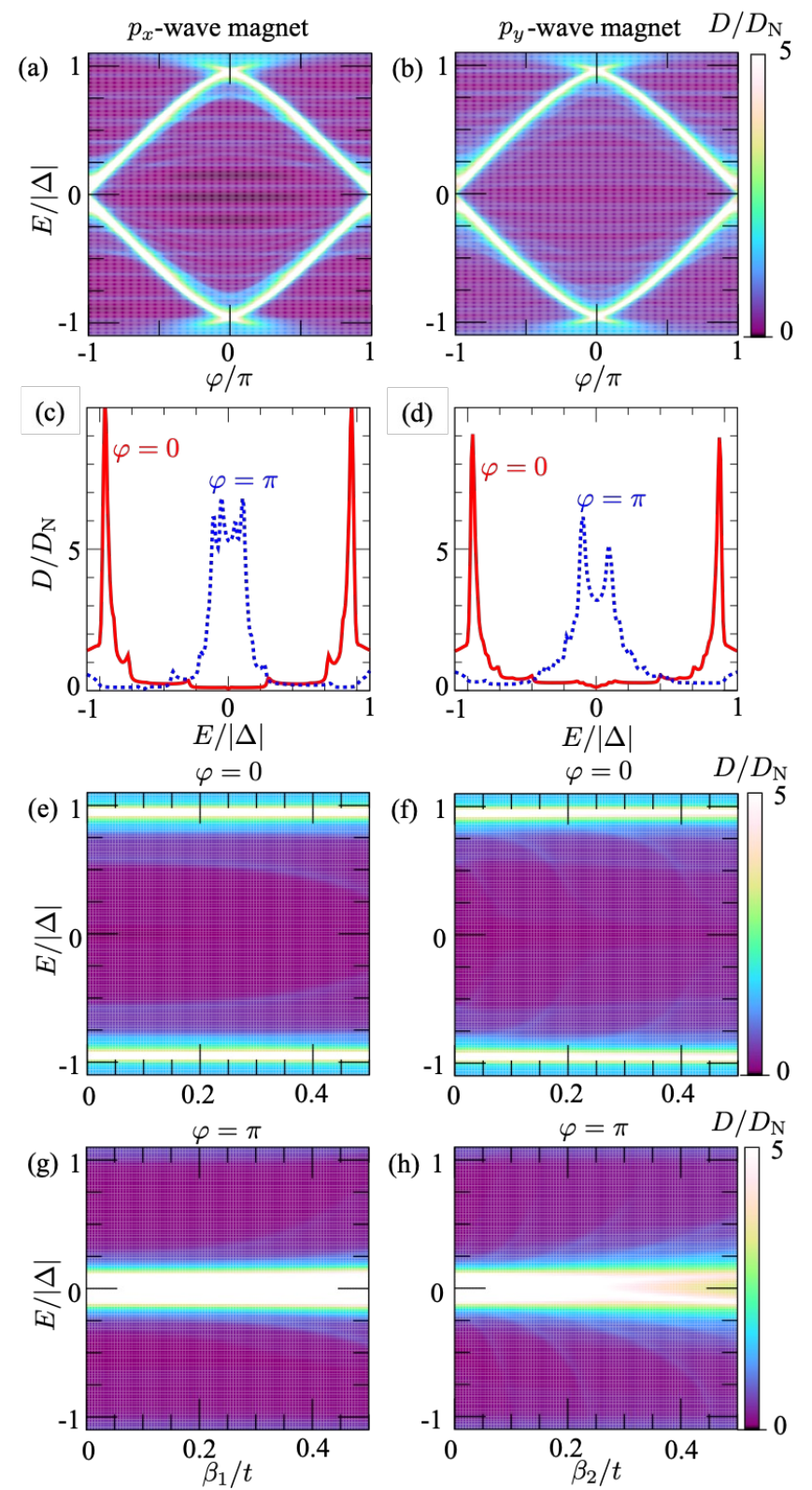}
    \caption{(a) and (b) LDOS normalized by its  zero-energy normal state value as a function of $\varphi$ and $E$ in the middle of a Josephson junction with $p_{x}$-wave and $p_{y}$-wave magnets. 
    (c) and (d) Line cuts of (a) and (b) as a function of $E$ at $\varphi=0,\pi$.
    (e) and (g) Normalized LDOS as a function of $E$ and strength of magnetic $p_{x}$-wave order $\beta_{1}$ at $\varphi=0,\pi$. (f) and (h) Same as (e) and (g) but as a function of $E$ and  strength of magnetic $p_{y}$-wave order $\beta_{2}$ at $\varphi=0,\pi$. Parameters: $(j,j')=(0,0)$, $\beta_{1}=0.5t$, $\beta_{2}=0.5t$, and the rest same as in Fig.~\ref{fig:4}.}
    \label{fig:5}
\end{figure}%

To detect the formation of ABSs discussed above, we now explore the angle-averaged LDOS in the center of the unconventional magnet using the first equation of Eqs.\,(\ref{LDOSI}). 
In Figs.~\ref{fig:4}(a) and \ref{fig:4}(b) and Figs.~\ref{fig:5}(a) and \ref{fig:4}(b), we present the LDOS $D$ normalized by the LDOS in the normal state at $E=0$ $D_{\rm N}$ as a function of the energy and phase difference $\varphi$ for the $d$-wave AM and UPM Josephson junctions discussed in the first part of this section. 
In Figs.~\ref{fig:4} (c) and \ref{fig:4}(d) and Figs.~\ref{fig:5}(c) and \ref{fig:5}(d) also present the LDOS as a function of $E$ at fixed $\varphi$ for $d$-wave AM and UPM Josephson junctions, while in Figs.~\ref{fig:4} (e) and \ref{fig:4}(f) and Figs.~\ref{fig:5}(e) and \ref{fig:5}(f) we show the LDOS as a function of $E$ and the strength of the unconventional magnet fields $\alpha_{1,2}$ and $\beta_{1,2}$.

From the LDOS as a function of energy $E$ and phase difference $\varphi$ in Figs.~\ref{fig:4}(a) and \ref{fig:4}(b), the first feature to notice is the appearance of high intensity regions that correspond to the ABSs for all configurations of $k_{y}$, see Fig.~\ref{fig:2}. 
While the LDOS for both the $d_{xy}$-wave and $d_{x^{2}-y^{2}}$-wave AM-based Josephson junctions exhibit signatures of ABSs, the visibility of the ABSs for the $d_{x^{2}-y^{2}}$-wave case is greater. 
Another interesting feature in the LDOS is that, for the $d_{xy}$-wave AM-based junction, it exhibits very small values near zero energy, revealed in the darkish region of Fig.~\ref{fig:4}(a); in contrast, the $d_{x^{2}-y^{2}}$-wave case has always sizable values. 
A closer inspection of the LDOS at fixed values of $\varphi$ shows that their small values for the $d_{xy}$-wave case form a V-shape gap structure near zero energy [Fig.~\ref{fig:4}(c)] irrespective of $\varphi$ but without vanishing at zero energy. Notably, for $d_{x^{2}-y^{2}}$-wave AM junction, the always finite LDOS can develop large values around zero energy due to zero-energy ABSs as it happens for $\varphi=0.8\pi$ in Fig.~\ref{fig:4}(d), see also Fig.~\ref{fig:2}(f). 
When varying the strength of the AM orders $\alpha_{1,2}$, the angle-resolved LDOS in the middle of the $d_{xy}$- and $d_{x^{2}-y^{2}}$-wave AM Josephson junctions develop interesting profiles for $\varphi=0,\pi$, see  Figs.~\ref{fig:4}(e)-\ref{fig:4}(h). Interestingly, for the $d_{xy}$-wave case at $\alpha_{1}=0$, the LDOS at $\varphi=0$  develops large peaks near the gap edges $E=\pm|\Delta|$, while at $\varphi=\pi$ it gets large values at zero energy $E=0$ [Figs.~\ref{fig:4}(e) and \ref{fig:4}(g)]; this is consistent with the behavior of conventional Josephson junctions without magnetic order. 
As $\alpha_{1}$ increases, the LDOS at $\varphi=0,\pi$ acquires an oscillatory behavior with peaks that occur around zero energy in an alternating fashion between $\varphi=0$ and $\varphi=\pi$ [Fig.~\ref{fig:4}(e)]. 
Although the periodicity of the oscillations is maintained as $\alpha_{1}$ increases, their amplitude gets smaller. 
In the case of the LDOS for $d_{x^{2}-y^{2}}$-wave AM Josephson junctions [Figs.~\ref{fig:4}(f) and \ref{fig:4}(h)], the LDOS has a similar start at $\alpha_{2}=0$, but, as  $\alpha_{2}$ takes finite values, the LDOS acquires large values at zero energy at $\varphi=0$ and $\varphi=\pi$ in an oscillatory way, see Figs.~\ref{fig:4}(f) and \ref{fig:4}(h). 
It is worth noting that the period of the oscillations for the LDOS in the $d_{x^{2}-y^{2}}$-wave case is larger than that for the $d_{xy}$-wave case, but the amplitudes of such oscillations decay at $\alpha_{1,2}$ increase. 
At the same time, the intensity of the LDOS becomes weaker at $\alpha_{1,2}$ increases, implying that the effective energy gap is also smaller.  

The angle-resolved LDOS for the $p_{x,y}$-wave magnet-based Josephson junctions also reveals the formation of ABSs [Figs.~\ref{fig:5}(a) and \ref{fig:5}(b)], but it is rather challenging to identify the spin split ABSs for the $p_{y}$-wave magnet seen in Fig.\,\ref{fig:3}(d).  
The LDOS in both cases is peaked at $E=\pm|\Delta|$ for $\varphi=0$, while at $\varphi=\pm\pi$ it develops large values near zero energy, see also Figs.~\ref{fig:5}(c) and \ref{fig:5}(d). 
Interestingly, the LDOS at $\varphi=\pm\pi$ develops a U-shape profile around zero energy, with peaks around zero energy and a finite value at zero energy, see Figs.~\ref{fig:5}(c) and \ref{fig:5}(d). 
Varying the system parameters, such as magnetic order or chemical potential, increases the zero-energy LDOS such that the U-shape profile becomes a wide zero-energy LDOS peak. 
For varying strengths of the magnetic order via $\beta_{1,2}$, the LDOS in the $p_{x,y}$-wave cases exhibit small changes that are not very visible. 
In a close inspection at $\varphi=0$, we note that the LDOS acquires smaller values around zero energy, featuring an effective energy gap that decreases as $\beta_{1,2}$ increase [Figs.~\ref{fig:5}(e) and \ref{fig:5}(f)]. 
On the contrary, the LDOS at $\varphi=\pi$ exhibits large values at zero energy, with roughly constant values around zero energy at small $\beta_{1,2}$ but with smaller zero-energy LDOS as $\beta_{1,2}$ increase that leads to a U-shape gap structure [Figs.~\ref{fig:5}(g) and \ref{fig:5}(h)]. 
Thus, the LDOS for UPM-based Josephson junctions reveal distinct features, which are purely associated with the nature of their ABSs.

 \begin{figure}[htbp]
    \centering
    \includegraphics[width=8.5cm]{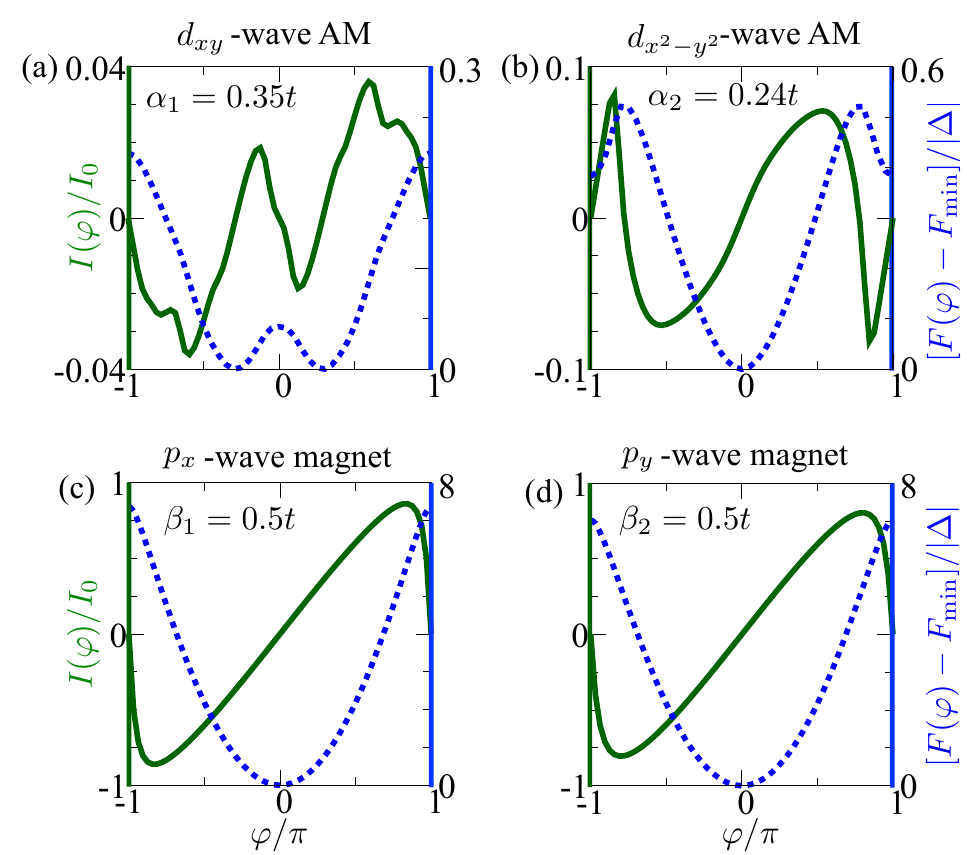}
    \caption{Josephson current  $I(\varphi)$ and the corresponding free energy \textcolor{black}{$F(\varphi)=\frac{\hbar}{2e}\int^{\pi}_{-\pi}I(\varphi)d\varphi$} as a function of the superconducting phase difference $\varphi$ for Josephson junctions with (a) $d_{xy}$- and (b) $d_{x^2-y^2}$-wave AMs, while (c) and (d) for $p_x$- and (d) $p_y$-wave magnets.
    \textcolor{black}{The green-solid and blue-dotted lines mean the Josephson current $I(\varphi)$ and the corresponding free energy $F(\varphi)-F_\mathrm{min}$, respectively.}
    The strength of the magnetic order is chosen as (a) $\alpha_{1}=0.35t$, (b) $\alpha_{2}=0.24t$, (c) $\beta_{1}=0.5t$, and (d) $\beta_{2}=0.5t$. Here, $I_{0}$ and \textcolor{black}{$F_\mathrm{min}=\min[F(\varphi)]$} stand for the critical current without the magnetic order 
    and the minimum value of the free energy, respectively.     
    Parameters: $\mu=1.5t$, $T=0.025T_\mathrm{c}$, $T_\mathrm{c}=0.01t$, $t_{\mathrm{int}}=0.95$, and $N=30$.}
    \label{fig:6}
\end{figure}%

\section{Current-phase curves and free energy}
\label{section4}

Having discussed the formation of ABSs, here we focus on the phase dependent Josephson current $I(\varphi)$ using Eqs.\,(\ref{LDOSI}) for Josephson junctions formed by $d$-wave AMs and UPMs, studied in Sect.~\ref{section3}. 
In Figs.~\ref{fig:6}(a) and \ref{fig:6}(b) we present the Josephson current $I(\varphi)$ \textcolor{black}{(green-solid line)} as a function of the superconducting phase difference $\varphi$ for Josephson junctions with $d_{xy}$- and $d_{x^{2}-y^{2}}$-wave AMs, while in Figs.~\ref{fig:6}(c) and \ref{fig:6}(d) for Josephson junctions with $p_{x}$- and $p_{y}$-wave magnets. 
In order to identify the type of Josephson junction, in Fig.~\ref{fig:6} we also present the respective free energy \textcolor{black}{$F(\varphi)=\frac{\hbar}{2e}\int^{\pi}_{-\pi} I(\varphi)d\varphi$} \textcolor{blue}{(blue-dotted line)} for each case, where we subtract the minimum value of the free energy \textcolor{black}{$F_{\rm min}=\min[F(\varphi)]$}.
In general, the Josephson current $I(\varphi)$ and the free energy $F(\varphi)$ in all cases are $2\pi$-periodic with $\varphi$ and vanish at $\varphi=0,\pm\pi$. 
However, they exhibit a distinct structure as $\varphi$ varies from $-\pi$ to $\pi$, which is strongly tied to the intrinsic nature of the unconventional magnetic order.

For Josephson junctions with $d_{xy}$-wave AMs in Fig.~\ref{fig:6}(a), $I(\varphi)$ is small and develops a highly unconventional phase dependence. 
In fact, within $0\leq\varphi\leq\pi$, we have that $I(\varphi)$ first takes negative values, develops a minimum value, and then acquires a positive maximum value, see a green curve in Fig.~\ref{fig:6}(a). 
Correspondingly, the free energy develops two minima at phases $\varphi<0$ and $\varphi>0$ and a maximum at $\varphi=0$, which coincide with the current taking zero values $I(\varphi)=0$. 
Because the free energy minima are not happening at zero phase, Fig.~\ref{fig:6}(a) reveals the emergence of a $\varphi$-Josephson junction entirely controlled by the AM order. 
We note that the $\phi$-Josephson junction obtained here is distinct to what occurs in  Josephson junctions with conventional SCs, which are $0$-Josephson junctions because the Josephson current behaves as $I(\varphi)\sim {\rm sin}(\varphi)$   and the free energy has a minimum at $\varphi=0$ \cite{Golubov2004}. 
For $d_{x^{2}-y^{2}}$-wave AM-based Josephson junctions, the current $I(\varphi)$ is also nontrivial and exhibits larger values than for $d_{xy}$-wave AM case, see green curve in Fig.~\ref{fig:6}(b). 
Here, for $0\leq\varphi\leq\pi$, the current first takes positive values and develops a maximum, which is followed by a sharp transition into negative values peaked at $\varphi\sim \pm 0.8\pi$. 
Interestingly, despite the unusual supercurrent behavior, the free energy exhibits maxima at $\varphi\sim \pm 0.8\pi$ and a minimum value at $\varphi=0$, which classifies these junctions as a $0$-Josephson junction, see blue curve in Fig.~\ref{fig:6}(b). 
In the case of $p_{x,y}$-wave magnet-based Josephson junctions, their current-phase curves exhibit a conventional sine-like behavior, with a fast sign change around $\varphi=\pm\pi$ that originates an almost sawtooth profile, see green curves in Figs.~\ref{fig:6}(c) and \ref{fig:6}(d).  
This supercurrent profile at  $\varphi=\pm\pi$ originates due to ABSs having a cosine-like phase dispersion and vanishing energy at $\varphi=\pm\pi$, see Fig.~\ref{fig:3}.  
The respective free energies of the $p_{x,y}$-wave cases develop a single minimum $\varphi=0$, implying that these junctions are $0$-Josephson junctions.
 
\begin{figure}[t!]
    \centering
    \includegraphics[width=8.5cm]{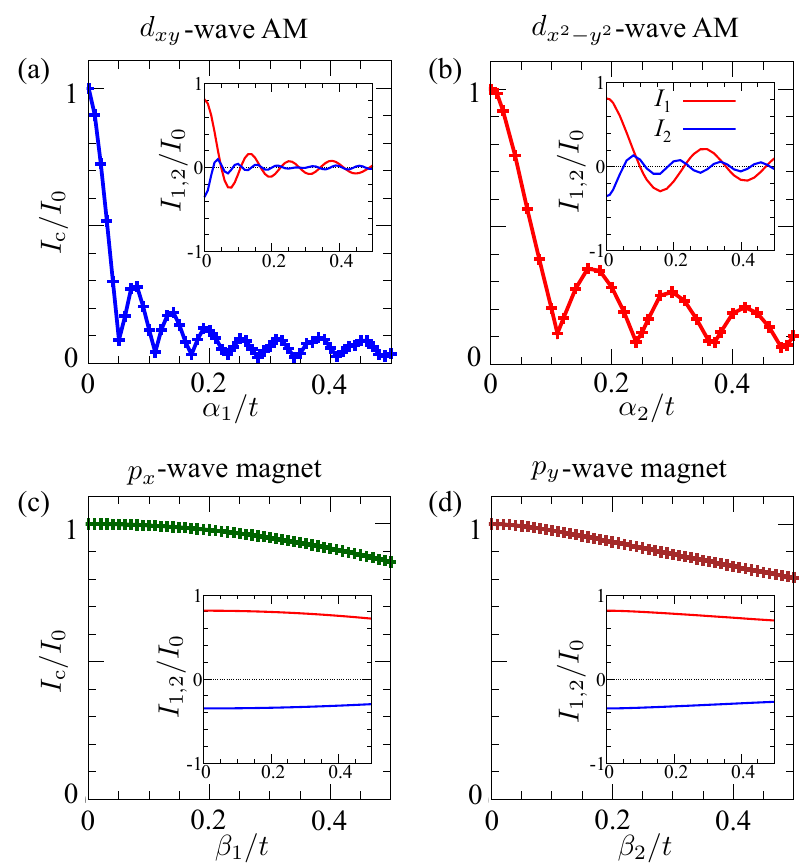}
    \caption{The critical current $I_\mathrm{c}$  as a function of the magnetic order strengths $\alpha_{1,2}$ and $\beta_{1,2}$ for the Josephson junctions based on  $d$-wave AMs (a) and (b), and UPMs (c) and (d), respectively. $I_{0}$ indicates the critical current without any unconventional magnetic order. The insets in (a-d) show the first and second harmonics $I_{1,2}$ of the supercurrent decomposed as in Eq.\,(\ref{eqIn2}). Parameters are the same as in Fig.~\ref{fig:6}.}
    \label{fig:7}
\end{figure}

To gain a further understanding of the Josephson currents, we now explore the maximum supercurrent $I_\mathrm{c}=\max[|I(\varphi)|]$, also known as critical current, where the maximum is taken over  $\varphi\in[-\pi,\pi]$. Moreover, to uncover the supercurrent structure shown in  Fig.~\ref{fig:6}, we exploit the periodicity of $I(\varphi)$ with respect to $\varphi$ and decompose it into series of harmonics as 
\begin{align}
\label{eqIn}
    I(\varphi)
    =\sum_{n}[I_{n}\sin(n\varphi)+J_{n}\cos(n\varphi)]\,,
\end{align}%
where $I_{n}$ and $J_{n}$ are the coefficients of the Fourier series, also referred to as $n$-th harmonics. 
While $I_{n}$ and $J_{n}$ are in general finite, the system symmetries can drastically affect their presence. 
In fact, the combined symmetries of the fourfold rotation operator $C_4$ and the time-reversal symmetry for $d$-wave AMs make the coefficients $J_{n}=0$ \cite{Bo2024}. 
Moreover, since the time-reversal symmetry is not broken in UPM, we get $J_{n}=0$ \cite{tanaka971}. 
Thus, the Josephson current $I(\varphi)$ in our setup can be decomposed as
\begin{align}
\label{eqIn2}
I(\varphi)=\sum_{n}I_{n}\sin(n\varphi)\,.
\end{align}%
Hence, we first obtain the Josephson current $I(\varphi)$   using the second expression of Eqs.\,(\ref{LDOSI}) and then decompose it using Eq.\,(\ref{eqIn2}) to identify the $n$-th harmonics contributing to $I(\varphi)$. In Fig.\,\ref{fig:7} we show the critical current $I_{c}$ as a function of the magnet order strength for the Josephson junctions with $d_{xy}$-wave and $d_{x^{2}-y^{2}}$-wave AMs as well as for $p_{x,y}$-wave magnets. Moreover, in the insets of Fig.\,\ref{fig:7} we also present the corresponding odd-parity first and second harmonics $I_{1,2}$ obtained by Fourier transforming Eq.\,(\ref{eqIn2}).  

The critical currents for the two types of considered $d$-wave AM-based Josephson junctions exhibit similar behavior and very slight differences, see Figs.\,\ref{fig:7}(a) and \ref{fig:7}(b). 
Among the similarities we find that the critical currents at $\alpha_{1,2}=0$ have an equal nonzero value which then undergoes a fast reduction developing a minimum within a small but finite $\alpha_{1,2}$, see Figs.\,\ref{fig:7}(a) and \ref{fig:7}(b). 
As $\alpha_{1,2}$ further increases, the critical currents develop an oscillatory decaying profile, with several minima where the critical currents reach vanishing values and a roughly constant periodicity. 
The oscillatory and vanishing critical currents can be understood from the profile of the harmonics that contribute to the Josephson current, which we obtain to be the first and the second harmonics $I_{1,2}$, as shown in the inset of Figs.\,\ref{fig:7}(a) and \ref{fig:7}(b). 
The sign of $I_{1,2}$ changes as $\alpha_{1,2}$ increases, indicating that the critical current oscillations are led by the $0$-$\pi$ transitions. 
The oscillatory decaying behavior of the critical currents as $\alpha_{1,2}$ increase is consistent with the decaying LDOS oscillations shown in Figs.\,\ref{fig:4}(e)-\ref{fig:4}(h), which stems from the behavior of the ABSs with $\alpha_{1,2}$ and are the ones reflected in the large intensity regions of the LDOS in Figs.\,\ref{fig:4}(e)-\ref{fig:4}(h). 
When $I_{1,2}$ change sign, they become equal, which thus explains the vanishing values of the critical currents, often referred to as critical current nodes. 
Interestingly, the separation between nodes, related to the periodicity in the critical currents, is distinct for $d_{xy}$-wave and  $d_{x^{2}-y^{2}}$-wave AM-based Josephson junctions, a signature that reflects the distinct $d$-wave altermagnetic order. 
It is also a signature of altermagnetism the fact that higher harmonic terms ${\rm sin}(n\varphi)$, with $n>0$, define the supercurrent profile \cite{Bo2024}.  
The oscillatory critical currents we show in Figs.\,\ref{fig:7}(a) and \ref{fig:7}(b) are akin to those found in topological Josephson junctions due to Majorana states \cite{cayao2018andreev,CayaoMajorana,PhysRevB.104.L020501,PhysRevB.105.054504,PhysRevB.109.L081405,baldo2023zero} but here topological superconductivity is absent. 

For $p_{x,y}$-wave magnet-based Josephson junctions, the critical currents exhibit a weak dependence on the strength of the UPM fields $\beta_{1,2}$, see Figs.\,\ref{fig:7}(c) and \ref{fig:7}(d).
First of all, the critical currents exhibit a very slow decay as $\beta_{1,2}$ increases but without oscillations, with an almost linear profile within the range of shown fields.
The corresponding first and second contributing harmonics $I_{1,2}$ here exhibit positive and negative values, with a larger value of $I_{1}$ in both $p_{x,y}$-wave magnets, see insets in Figs.\,\ref{fig:7}(c) and \ref{fig:7}(d). 
While the second harmonic $I_{2}$ shows a tiny enhancement as $\beta_{1,2}$ increases, the first harmonic $I_{1}$ develops a notorious decay [Figs.\,\ref{fig:7}(c) and \ref{fig:7}(d)], albeit smaller when compared with the $d$-wave AM-based Josephson junctions [Figs.\,\ref{fig:7}(a) and \ref{fig:7}(b)]. 
The weak dependence of the critical currents on the UPM fields $\beta_{1,2}$ and therefore serves as a useful indicator for distinguishing between $d$-wave altermagnetism and $p$-wave magnetism in Josephson junctions.

 \begin{table*}[t!]
    \centering
    \begin{tabular}{c|c|c|c}
        Magnetic order & Parity of $M_{zx}$ & Emergent pair symmetry $[k_y\leftrightarrow -k_{y},j\leftrightarrow j']$ \\ \hline
        Normal metal &  Even& ESE$[+,+]$+OSO$[+,-]$ \\
        Ferromagnet &  Even&  ESE$[+,+]$+ETO$[+,-]$+OTE$[+,+]$+OSO$[+,-]$ \\
        $d_{xy}$-wave AM & Odd & ESE$[+,+]$+ETO$[-,+]$+OTE$[-,-]$+OSO$[+,-]$ \\
        $d_{x^2-y^2}$-wave AM  & Even  & ESE$[+,+]$+ETO$[+,-]$+OTE$[+,+]$+OSO$[+,-]$ \\
        $p_{x}$-wave magnet & Even  & ESE$[+,+]$+ETO$[+,-]$+OTE$[+,+]$+OSO$[+,-]$ \\
        $p_{y}$-wave magnet & Odd  & ESE$[+,+]$+ETO$[-,+]$+OTE$[-,-]$+OSO$[+,-]$ \\
    \end{tabular}
    \caption{Pair symmetries in Josephson junctions formed by spin-singlet $s$-wave pairing SCs and unconventional magnets (AMs and UPMs) emerging under the exchange of frequency $i\varepsilon_{n}$, spins $(\sigma,\sigma')$, momentum $k_{y}$ and spatial coordinates along $x$ inside the magnets $(j,j')$. 
    The first and second rows correspond to the pair symmetries when the unconventional magnet is replaced by a normal metal or ferromagnet, respectively. 
    The first column indicates the type of magnetic system, while the second column represents the parity of the mirror operator $M_{zx}$ for the respective magnetic orders in the $zx$-plane. 
    The third column shows the allowed pair symmetries ESE, ETO, OTE, and OSO, where the first, second, and third letters in this notation correspond to the pair amplitude being even (or odd) under the exchange of frequency $(i\varepsilon_{n}\leftrightarrow-i\varepsilon_{n})$, spin $(\sigma\leftrightarrow\sigma')$, and spatial coordinates $(j\leftrightarrow j')$ plus momentum $(k_y\leftrightarrow -k_{y})$. 
    The $+/-$ signs in square brackets indicate the evenness/oddness of the pair amplitude under the exchange of momentum $k_{y}$ and spatial coordinates, namely, $[k_y\leftrightarrow -k_{y},j\leftrightarrow j']$. 
    Note that locally in space $j=j'$, the odd parity state under exchange of spatial coordinates is zero, hence OSO vanishes; OSO is a pair symmetry fully nonlocal in space and exists irrespective of the magnetic order. 
    The triplet states, however, require magnetism.}
    \label{tab:1}
\end{table*}%

The critical currents for Josephson junctions with $d$-wave AMs and UPMs (Fig.\,\ref{fig:7}) can be also intuitively understood from the sequential Andreev reflections occurring at both interfaces between SCs and unconventional magnet \cite{Kashiwaya_2000}. 
This is because Andreev reflections are the fundamental scattering processes determining the Josephson current, which involves the transfer of Cooper pairs between SCs via the unconventional magnet. 
Thus, for $d$-wave AM bases Josephson junctions, given that the considered SCs host spin-singlet $s$-wave Cooper pairs and that the AM fields $\alpha_{1,2}$ produce spin-polarized bands, the Andreev reflections are suppressed by the increase of $\alpha_{1,2}$.
\textcolor{black}{Then the Cooper pair feels the variety of the net magnetization for each transverse momentum $k_y$ and $\alpha_{1,2}$ even though total magnetization is zero in AMs.}
This effect thus reduces the transfer of Cooper pairs and therefore causes a rapid decrease of critical currents $I_c$ in $d$-wave AM bases Josephson junctions, as observed in Figs.\,\ref{fig:7}(a) and \ref{fig:7}(b). 
This mechanism also explains the $0$-$\pi$ transitions associated with the critical current oscillations in $d$-wave AM-based Josephson junctions \cite{zhang2024}: this is because Cooper pairs acquire an oscillating factor at fixed $k_{y}$ proportional to $e^{i (k_{x\uparrow }-k_{x\downarrow }) x}$, where $k_{x\uparrow}$ and $k_{x\downarrow}$ are the wavevectors of the paired electrons which are usually different in the middle $d$-wave AM region.   
\textcolor{black}{We note that the potential barrier at the interface also leads to a drastic change in the feature of the junctions, e.g.\ 0-$\pi$ phase transition, as worked in Ref.\ \cite{Bo2024}.}
In the case of Josephson junctions with UPMs [Figs.\,\ref{fig:7}(c) and \ref{fig:7}(d)], the time-reversal symmetry is present, and hence the paired electrons have opposite spin directions. 
In consequence, the Andreev reflections are more likely to occur and the decrease of critical current in UPM-based Josephson junctions becomes much slower than in AMs. 
Moreover, since $k_{x\uparrow }=k_{x\downarrow }$ for the pairing in UPM, there is no oscillatory factor in the Cooper pairing, explaining the absence of $0$-$\pi$ oscillations in Josephson junctions [Figs.\,\ref{fig:7}(c) and \ref{fig:7}(d)].
The decaying profile of $I_c$ for the UPM-based Josephson junctions is mainly caused by the Fermi surface mismatch. 
In fact, for $p_{y}$-wave magnets, states with $\left\vert k_{y}\right\vert >k_{F}$, with $k_{F}$ is the Fermi wavevector of the SC, are unable to anticipate the transport process because the number of states increases as $\beta_{2}$ increases, thus resulting in the suppression of the critical current. 
In contrast, for  $p_{x}$-wave magnets, the splitting of the Fermi surface is along the $k_{x}$-direction, and hence most of the states obey $\left\vert k_{y}\right\vert \leq k_{F}$, which explains why current decays very slowly.  
This discussion therefore explains why $I_c$ decays faster in Josephson junctions with $p_{y}$-wave magnet than with $p_{x}$-wave one.

\section{Emergent odd-frequency pairing}
\label{section5}
Since the Josephson current studied here involves the transfer of Cooper pairs between SCs via unconventional magnets, in this part we study the types of Cooper pairs that emerge in the unconventional magnets and hence allow the entire Josephson effect. 
To identify the type of emergent Cooper pairs, we explore the symmetries of the superconducting correlations in the middle of the unconventional magnet. 
For this purpose, we inspect the anomalous electron-hole Green's functions, known also as pair amplitudes, $F_{j,j'}^{\sigma,\sigma'}(k_{y},i\varepsilon_{n})$, where $(j,j')$ denote position coordinates inside the AM, $k_{y}$ is the wavevector along $y$ being a good quantum number, $(\sigma,\sigma')$ the paired electron spins, and $i\varepsilon_{n}$ are   Matsubara frequencies.
We note that $F_{j,j'}^{\sigma,\sigma'}(k_{y},i\varepsilon_{n})$ corresponds to the off-diagonal elements of the Green's functions obtained in Eq.\,(\ref{GF}) of Sect.~\ref{section2}. 
Now, to identify the symmetries of $F_{j,j'}^{\sigma,\sigma'}(k_{y},i\varepsilon_{n})$, we exploit the fact that it represents paired electrons, implying that $F_{j,j'}^{\sigma,\sigma'}(k_{y},i\varepsilon_{n})$ must be antisymmetric under the exchange of all quantum numbers, namely, $(j,j')$, $k_{y}$, $(\sigma,\sigma')$, and $i\varepsilon_{n}$. 
Thus, the pair amplitude obeys
\begin{equation}
\label{Fantisymmetry}
F_{j,j'}^{\sigma,\sigma'}(k_{y},i\varepsilon_{n})=-F_{j',j}^{\sigma',\sigma}(-k_{y},-i\varepsilon_{n})\,.
\end{equation}
By taking into account this antisymmetry condition, we can identify all the allowed combinations of quantum numbers. Under the individual exchange of quantum numbers, e.g., $j\leftrightarrow j'$, the pair amplitude can be an \textit{even} or \textit{odd} function with respect to the quantum numbers.  
The same applied for the exchange with respect to frequency $(i\varepsilon_{n}\leftrightarrow-i\varepsilon_{n})$, spins $(\sigma\leftrightarrow\sigma')$, and momentum $k_{y}\leftrightarrow -k_{y}$. 
Taking all these possibilities, we find that there are four pair symmetry classes allowed: even-frequency, spin-singlet, even-parity (ESE); even-frequency, spin-triplet, odd-parity (ETO); odd-frequency, spin-triplet, even-parity (OTE), and odd-frequency, spin-singlet, odd-parity (OSO) states~\cite{bezrezinskij1974,
Balatsky1992,odd1,odd3,Eschrig2007,Tanaka2012,Fominov,LinderRev19,Cayao2020,triola2020role,Tanaka2024}. 
The \textit{parity} symmetry involves the exchange in spatial coordinates $(j,j')$ and the exchange in momentum $k_{y}$; hence the distinct parity symmetries are induced by the superconducting interfaces and also by the anisotropy of the magnetic order along $y$. 
Also, the allowed pair symmetries involve spin triplet which are expected due to the unconventional magnetic order of the considered AMs and UPMs: given that the parent SCs have ESE symmetry, the unconventional magnetic order drives a spin-singlet to spin-triplet conversion. 
Since the parent SCs have ESE symmetry and the spin direction of the unconventional magnets is along $z$, only mixed spin triplet is allowed where the paired electrons have opposite spin~\cite{Eschrig2007,Yokoyama2007,Efetov2,Tanaka2012}. 
All the possible pair symmetries in Josephson junctions with AMs and UPMs are summarized in   Table \ref{tab:1}, where, for completeness, we also add the possible pair symmetries when the mediating region between SCs is a normal metal or a ferromagnet.

Under general circumstances, the four pair symmetry classes are allowed in Josephson junctions with AMs and UPMs. 
They, however, depend on the nature of the magnetic order and on the parity. 
For instance, when the mirror symmetry of the magnetic order is even with respect to the $zx$-plane, such as in a ferromagnet, $d_{x^2-y^2}$-wave AM, or $p_x$-wave magnet, the induced spin-triplet pair correlations have ETO and OTE symmetries: these pair symmetries are, respectively, odd and even under the exchange of the spatial coordinates $(j,j')$, see second, fourth, and fifth rows in Table \ref{tab:1}.  
This implies that ETO $p_x$-wave and OTE $s$- or $d_{x^2-y^2}$-wave pairings are induced in the UPM and AM, respectively. 
With spin singlet, OSO pair symmetry is induced which is odd under the exchange of $(j,j')$, but this state also exists even without unconventional magnetism, see the first row of Table \ref{tab:1}. 
For $d_{xy}$-wave AM and $p_y$-wave magnet having a mirror symmetry that is odd with respect to the $zx$-plane, the induced pair symmetries are ETO and OTE  which have even and odd under the exchange of spatial coordinates but are odd and even under momentum exchange $k_{y}\leftrightarrow -k_{y}$.  
Hence, ETO $p_y$-wave and OTE $d_{xy}$-wave pairings emerge in the UPM and AM, respectively. We note that the OTE $d_{xy}$-wave pairing for the mirror-odd magnet originates from the opposite magnetization at $k_y$ and $-k_y$. 
We can therefore conclude that the unconventional magnet with even (odd) mirror symmetry does not change (can change) the total parity of the emergent superconducting pairing. 
Nevertheless, the unconventional magnet promotes mixed spin-triplet pairs which then assist the transfer of Cooper pairs between SCs so that the Josephson current gets affected.

\begin{figure}[t!]
    \centering
    \includegraphics[width=8.5cm]{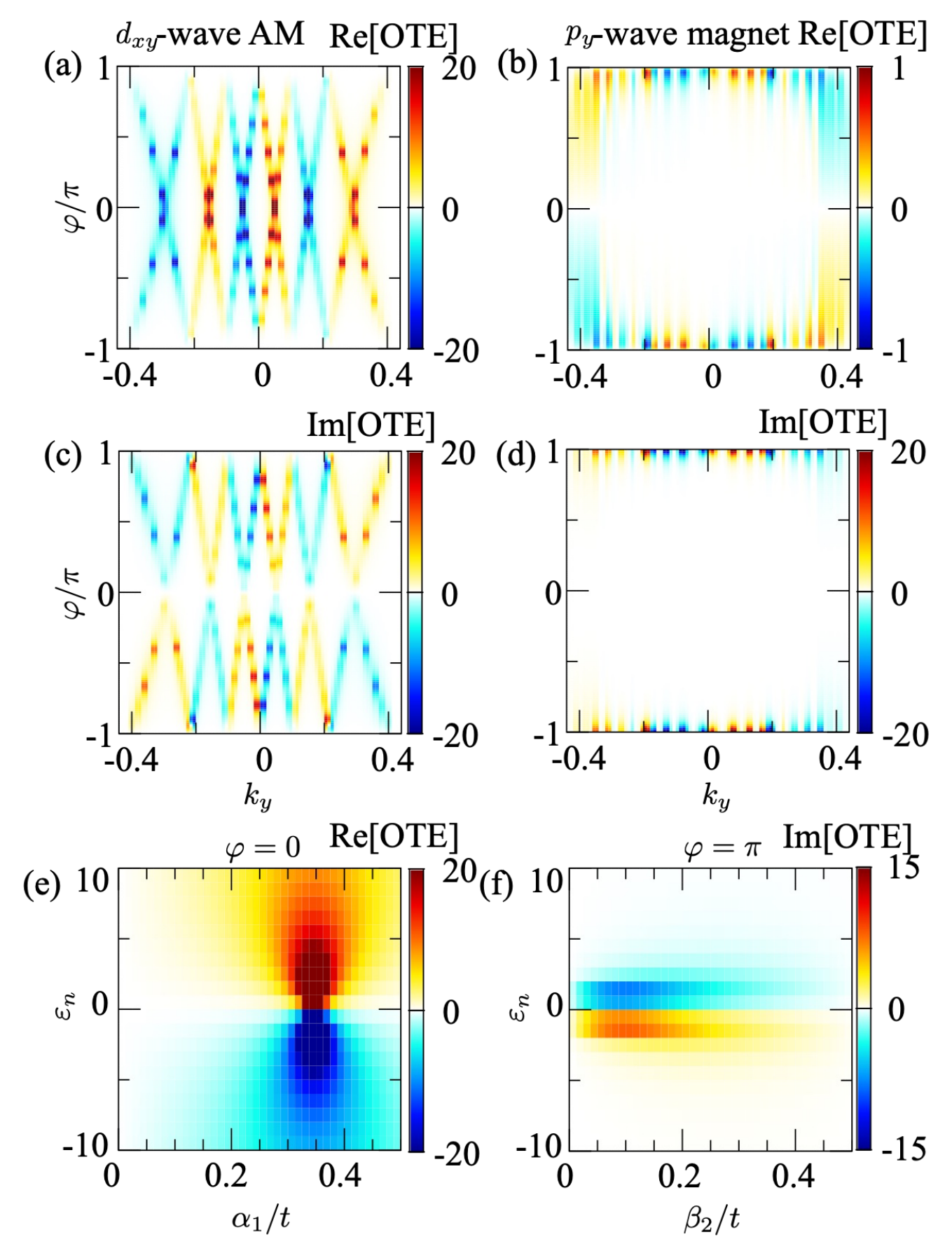}
    \caption{(a) and (c) Real (Re) and imaginary (Im) OTE pair amplitudes in the middle of a  $d_{xy}$-wave AM in a Josephson junction as a function of $\varphi$ and $k_{y}$, while (b) and (d)  in the middle of the $p_{y}$-wave magnet.  
    In (a)-(d), the Matsubara frequency is fixed at $i\varepsilon_n=i\pi T$.
    (e) and (f) Same quantities as in (a)-(d) as a function of Matsubara frequency $\varepsilon_n$ and strength of magnetic order$\alpha_{1}$ and $\beta_{2}$ at $\varphi=0,\pi$.   
    Parameters: $\alpha_{1}=0.35t$, $\beta_{2}=0.5t$, $k_{y}=0.05\pi$,
   $\mu=1.5t$, $T=0.025T_\mathrm{c}$, $T_\mathrm{c}=0.01t$, $t_\mathrm{int}=0.95$, and $N=30$.}
    \label{fig:8}
\end{figure}%

To visualize the above discussion, in Figs.\,\ref{fig:8}(a)-\ref{fig:8}(d) we present the real (Re) and imaginary (Im) OTE pair amplitudes in the middle of a $d_{xy}$-wave and $p_{y}$-wave magnet of a Josephson junction as a function of $\varphi$ and $k_{y}$ at $i\varepsilon_n=\pm i\pi T$, $\alpha_{1}=0.35t$, $\beta_{2}=0.5t$. 
In Figs.\,\ref{fig:8}(e) and \ref{fig:8}(f) we also show the Re and Im parts of the OTE amplitudes as a function of $\varphi$ and $\alpha_{1}$ and $\beta_{2}$ at $\varphi=0,\pi$, $k_{y}=0.05\pi$. 
We note that the plotted OTE pairing is odd under the exchange of spatial coordinates $(j,j')$ and odd under momentum $(k_{y})$, resulting in a total parity that is even. 
For the $d_{xy}$-wave AM in Figs.\,\ref{fig:8}(a) and \ref{fig:8}(b), both the Re and the Im parts of OTE oscillate as a function of $k_{y}$, while variations of $\varphi$ reflect the emergence of phase-dependent ABSs. 
While the Re OTE pairing exhibits large values at zero phase $\varphi=0$, its Im part has vanishing values at $\varphi=0$. 
At fixed values of momentum and phase, $k_{y}=0.05$ and $\varphi=0$ in Fig.\,\ref{fig:8}(e), the OTE pairing develops the expected odd-frequency dependence, changing sign around zero energies. 
Interestingly, at $\alpha\sim0.35t$, the OTE pairing acquires large values around zero energy, which correspond to the zero-energy ABSs shown in Fig.\,\ref{fig:2}. 
The appearance of large OTE pairing near zero energy and its connection with zero-energy ABSs has also been shown before but in Josephson junctions with unconventional SCs and Majorana states \cite{Proximityp,odd2,odd3,Tanaka2012,Cayao2020,Tanaka2024}; see also Refs.\,
\cite{PhysRevB.87.104513,Lu_2015,PhysRevB.92.100507,PhysRevB.92.205424,PhysRevB.95.174516,PhysRevB.95.184506,PhysRevB.96.155426,thanos2019,Takagi18,parhizgar2020large,PhysRevB.97.075408,PhysRevB.97.134523,PhysRevB.101.094506,PhysRevB.101.214507,PhysRevB.106.L100502,PalPRB2024,PAL2025116127}. 
In the case of Josephson junctions with $p_{y}$-wave magnet [Figs.\,\ref{fig:8}(b),\ref{fig:8}(d), and \ref{fig:8}(f)], the induced OTE pairing has the same parity as in the $d_{xy}$-wave AM case but exhibits an overall distinct profile. 
The most evident feature is perhaps its vanishing values around $\varphi=0$ and $k_{y}=0$ at finite frequencies, as seen in Figs.\,\ref{fig:8}(b) and \ref{fig:8}(d). 
Close to $\varphi=\pm\pi$, however, the OTE pairing acquires a finite oscillatory profile as $k_{y}$ is varied, revealing that both the Re and Im parts are odd functions of $k_{y}$, while only the Re part is an odd function of $\varphi$; this behavior is distinct to what we obtain for the OTE pairing in the $d_{xy}$-wave AM in Figs.\,\ref{fig:8}(a) and \ref{fig:8}(c). 
At fixed momentum and phase ($k_{y}=0.05\pi$ and $\varphi=\pi$) in Fig.\,\ref{fig:8}(f), the OTE pairing as a function of frequency and UPM magnetic order $\beta_{2}$ changes sign as $i\varepsilon_{n}$ changes from positive to negative values, thus revealing its odd frequency symmetry. 
Figure\ref{fig:8}(f) also shows that around zero frequency, OTE gets enhanced over a large range of both $\beta_{2}$ and frequencies, which can be seen similar to what occurs in $d_{xy}$-wave AM at $\varphi=0$   [Fig.\,\ref{fig:8}(e)]. 
However, the OTE pairing in the $p_{y}$-wave magnet case achieves large values at $\varphi=\pi$ and at $\beta_{2}\sim0.1\pi$, which is smaller than the needed magnetic strength in $d_{xy}$-wave AMs. 
The enhancement of the OTE pairing in $p_{y}$-wave magnet-based Josephson junctions is also attributed to the emergence of zero-energy ABSs at $\varphi=\pi$.

\section{Conclusions}
\label{section6}
{In conclusion, we have investigated the emergence of   Andreev bound states and their impact on phase-biased supercurrents and emergent superconducting correlations in Josephson junctions formed by spin-singlet $s$-wave superconductors and unconventional magnets. Specially, we focused on $d$-wave altermagnets with $d_{xy}$- and $d_{x^{2}-y^{2}}$-wave symmetries and on magnets with $p_{x,y}$-wave symmetry. We have found that the Andreev bound states in junctions with $d$-wave AMs and $p_y$-wave magnets strongly respond to variations of the transverse momentum, while they are not affected in $p_x$-wave magnet-based junctions. Of relevance here is that the transverse momentum dependence allows Andreev bound states to form near zero energy at distinct values of the superconducting phase difference $\varphi$. We have then shown that the signatures of the Andreev bound states can be detected via the local density of states, including their oscillatory decaying dependence as the strength of the unconventional magnetic order increases. We have then shown that the Josephson current in junctions with $d$-wave altermagnets develops an anomalous dependence on $\varphi$, which includes the contribution from higher harmonics and even the appearance of $\phi$-junction behavior in the $d_{xy}$-wave case. We have further demonstrated that the critical currents in junctions with  $d$-wave altermagnets develop an oscillatory decaying profile as a function of the field strength, while such oscillations are absent for $p$-wave magnet junctions but a slow decay persists. We finally uncovered that the combination of unconventional magnetism and the Josephson effect induces controllable odd-frequency spin-triplet $s$-wave  pair correlations in the unconventional magnets, reflecting the nature of the proximity effect in these materials.  

\textcolor{black}{We discuss the role of the magnetic anisotropy in Josephson junctions.
We can assume two types of magnetic anisotropy: (1) the direction of the magnetization and (2) the rotation
of the exchange energy at the $z$-axis in $\hat{H}_\mathrm{N}(\bm{k})$.
In case (1), because we consider the spin-singlet $s$-wave pairing in the present study, novel phenomena, i.e.\ the superconducting diode effect~\cite{FAndoNature2020}, are not expected.
In case (2), when we assume the arbitrary direction of the magnetic anisotropy for the interface, we expect strong momentum-dependent ABSs for the $d$-wave AM, see also in Ref.\ \cite{Bo2024}.
Thus, the Josephson current is also oscillatory decaying as a function of the altermagnetic order shown in Ref.\ \cite{Bo2024}
For the UPM order, we expect momentum-dependent ABSs, however, its dependence is weaker than that for $d$-wave AMs. 
Hence, we do not expect a dramatic change in the results in Josephson junctions with UPMs. 
In the odd-frequency pair amplitude in junctions, the odd-frequency/spin-triplet/even-parity state in which $[+,+]$ and $[-,-]$ are mixed because the mirror symmetry of the exchange field in the horizontal plane is broken in the arbitrary direction of the magnetic anisotropy.
}

Given the recent advances, hybrid junctions based on conventional superconductors and unconventional magnets as the ones studied are within experimental reach. Moreover,  the signatures of Andreev bound states in the local density of states can be in principle detected by scanning tunneling
spectroscopy \cite{Binnig87,Fischer07} and also by conductance \cite{Kashiwaya_2000,Mondal24}, while the phase-biased Josephson transport can be implemented following the spirit of superconductor-semiconductor systems~\cite{Lutchyn2018,prada2019andreev}. Furthermore, the existence of a finite zero energy peak in the density of states accompanied by odd-frequency spin-triplet $s$-wave pairing reveals the manifestation of what is known as the anomalous proximity effect \cite{Proximityp,Proximityp2,Proximityp,odd1,Yokoyama2007,Lin2021,Lin2023}.
Our work provides a fundamental understanding of proximity induced superconductivity and the Josephson effect in unconventional magnets, which can be useful for designing future superconducting spintronics devices.
}

\textcolor{black}{Although we do not focus on Josephson junctions with an antiferromagnet (AFM), we comment on the possible Josephson current in Josephson junctions with an AFM.
In previous studies~\cite{PhysRevLett.96.117005,PhysRevB.88.214512,PhysRevResearch.1.033095}, 0-$\pi$ phase transition was proposed in junctions based on the antiferromagnetic layers.
In Refs.\ \cite{PhysRevB.72.184510,PhysRevB.95.104513}, ABSs and LDOS were shown in superconductor/AFM/superconductor junctions.
Because the total magnetization is zero in AFM, Josephson effect in superconductor/AFM/superconductor junctions can be understood by the contributions of the ferromagnetic order $m_z$ and $-m_z$ in superconductor/ferromagnet/superconductor junctions.
The behavior of the Josephson current at a ferromagnetic order $m_z$ is the same as that at $-m_z$ in spin-singlet $s$-wave superconductors Josephson junctions.
In addition, when we compare with the case of AMs, we can focus on the orientation dependence of the Josephson current.
The Josephson effect with AMs has a strong orientation dependence and it means that the feature of the Josephson current greatly changes if we change the orientation of the axis of AM from $d_{xy}$-wave AM to $d_{x^2-y^2}$-wave one~\cite{Bo2024}.
On the other hand, because we expect that AFM is not orientation-dependent, we predict that the Josephson effect does not depend on the orientation of the crystal axis of AFM. 
Hence, a similar behavior in superconductor/AFM/superconductor junctions can be expected in superconductor/ferromagnet/superconductor junctions.
}


\section{Acknowledgements}
Y.\ F.\ acknowledges Okayama University for the numerical calculation support.
K.\ M.\ thanks S.\ Ikegaya for valuable discussions. 
K.\ Y.\ acknowledges the support from Sumitomo Foundation. 
J.\ C.\ acknowledges financial support from the Swedish Research Council  (Vetenskapsr\aa det Grant No.~2021-04121) and the Carl Trygger’s Foundation (Grant No. 22: 2093). 
Y.\ T.\ acknowledges financial support from JSPS with Grants-in-Aid
for Scientific Research (KAKENHI Grants Nos.\ 23K17668, 24K00583, and 24K00556). B.\ L.\ is supported by the National Natural Science Foundation of China (project 12474049). 
We thank J.\ Liu for the valuable comments.


\bibliography{biblio}

\end{document}